\documentclass[preprint,superscriptaddress,amsmath,amssymb,aps,physrev,showkeys]{revtex4-2}
\usepackage{graphicx}
\usepackage{dcolumn}
\usepackage{bm}
\usepackage{newtxtext,newtxmath}
\usepackage{hyperref}
\hypersetup{
     colorlinks = true,
     linkcolor = blue,
     anchorcolor = blue,
     citecolor = blue,
     filecolor = blue,
     urlcolor = blue
     }

\usepackage[normalem]{ulem}
\usepackage[mathlines]{lineno}

\begin{document}
\title{Anisotropic medium-range order uncovers dynamic crossovers in glass-forming liquids}

\author{Kamlesh Mishra}
\affiliation{Chemistry and Physics of Materials Unit, Jawaharlal Nehru Centre for Advanced Scientific Research, Bangalore 560064, India
}
\author{Rajesh Ganapathy}%
\affiliation{Chemistry and Physics of Materials Unit, Jawaharlal Nehru Centre for Advanced Scientific Research, Bangalore 560064, India
}
\affiliation{International Centre for Materials Science, Jawaharlal Nehru Centre for Advanced Scientific Research, Bangalore 560064, India}
\affiliation{School of Advanced Materials (SAMat), Jawaharlal Nehru Centre for Advanced Scientific Research, Bangalore 560064, India}
\author{Walter Kob}
\email{walter.kob@umontpellier.fr}
\affiliation{Department of Physics, University of Montpellier, CNRS, Montpellier F-34095, France}

\date{\today}

\begin{abstract}
On cooling liquids towards their glass-transition temperature, the dramatic increase of their relaxation times is accompanied by changes in particle dynamics at two distinct temperatures: A high-$T$ crossover, where particles become temporarily caged by their neighbors, and a low-$T$ crossover, where the cage escape mechanism changes.
While there is some evidence that the former is associated with a change in local particle arrangement, no structural modification has so far been detected across the low-$T$ crossover, fueling scepticism about the relevance of structure for glassy dynamics.
Here, we introduce a novel four-point correlation function which allows to determine a structural length scale characterizing cage anisotropy.  
Extensive molecular dynamics simulations reveal that this scale, as well as the mean structural length scale of the glass-former, extends into the medium range, i.e., significantly exceeds the particle size. Strikingly, the difference between these two scales - a measure of the degree of cage anisotropy - peaks at both crossover temperatures. We discuss how the presence of these peaks enables understanding the nature of the change in the microscopic transport mechanism at the two temperatures, thereby linking both dynamical crossovers to a single
structural observable, the anisotropic medium-range order. This fundamental insight demonstrates that structure, extending well-beyond the local cage, is an essential ingredient for understanding the relaxation dynamics of deeply supercooled liquids.
\end{abstract}

\maketitle

\section{Introduction}
The evolution of liquids into glasses upon cooling is marked by two significant qualitative changes in the particle dynamics~\cite{sastry1998signatures, andersen2005molecular, berthier2011theoretical, binder2011glassy, chandler2010dynamics}. The first occurs at the onset temperature, $T_o$, where particles become temporarily trapped by their neighbors in transient cages, thus signaling the beginning of glassy dynamics. This caging manifests as a plateau at intermediate times in observables like the mean-squared displacement or the intermediate scattering function. Previous studies have shown, that for certain glass-formers the local structure starts to change at $T_o$~\cite{malins2013identification,coslovich2011locally}, but no generic signature of this change has been identified so far. For temperatures below $T_o$, structural relaxation starts to slow down quickly,  becomes increasingly cooperative and the liquid starts to fragment into regions of high and low particle mobilities that fluctuate in space and time~\cite{ediger2000spatially}.
These dynamical heterogeneities (DH) are a universal feature of glass-forming systems and key to understanding how relaxation occurs, though a definitive explanation for their emergence is still lacking~\cite{berthier2011dynamical}.

A second important change in the dynamics occurs with further supercooling. In an Arrhenius representation, the structural relaxation time, $\tau_\alpha$, exhibits a distinct upwards bend at a temperature $T_X$~\cite{gotze1999recent, andersen2005molecular, berthier2011theoretical, binder2011glassy, lubchenko2007theory, cavagna2009supercooled}. Note that this second crossover is not only observed in fragile glass-formers, but also in strong ones~\cite{rossler1998universal,mallamace2010transport}, i.e., it is a universal feature of glass-forming liquids.
The mode-coupling theory (MCT) of the glass-transition predicts the existence of a temperature $T_c$ at which the $T-$dependence of $\tau_\alpha$ changes rapidly~\cite{gotze2009complex} and hence this ``critical temperature'' is commonly identified with $T_X$. A multitude of experiments and simulations have given ample evidence that MCT is able to make accurate prediction on the relaxation dynamics of glass-forming liquids close to $T_c$ (von Schweidler law, factorization property, wave-vector dependence of the non-ergodicity parameter,...)~\cite{gotze2009complex}, supporting the relevance of this theoretical approach. In addition, simulations and analytical calculations have found that for $T\lesssim T_c$, saddles in the potential energy landscape start to become significant obstacles that need to be crossed by the particles~\cite{angelani2000saddles, cavagna2001role, coslovich2006there, coslovich2019localization}, thus providing a more microscopic view of the dynamics around $T_c$. Furthermore, $T_c$ has also been found to correspond to a change in morphology of DH from string-like to compact clusters~\cite{stevenson2006shapes, kob2012non,hima2015direct, ortlieb2023probing}. Collectively, these findings suggest a crossover to a different transport mechanism near $T_c$. 

Despite these crucial insights, the mechanism giving rise to the observed upturn in $\tau_\alpha(T)$ around $T_c$ remains contentious. The main reason fueling this disagreement is that so far it has not been possible to identify any structural observable that signals the presence of $T_c$. Closely related to this point is the long-standing question regarding the nature of the glass transition: Is the slowing down of the dynamics a purely kinetic phenomenon~\cite{stillinger2013glass, biroli2013perspective,chandler2010dynamics}, or is it due to a subtle structural change~\cite{flenner2006hybrid, coslovich2018dynamic, das2022crossover}. To date, only dynamical length scales have shown a change near $T_c$ ~\cite{kob2012non, hima2015direct}. This raises the question whether this temperature has a real physical meaning in terms of structure or is only relevant for dynamical quantities. 

Here we employ a generalization of a recently proposed four-point correlation function~\cite{zhang2020revealing} to characterize the three-dimensional (3D) structure of disordered systems at distances well beyond the local cage~\cite{yuan2021connecting, singh2023intermediate, zhang2024revealing, zhang2025symmetry, yuan2025amorphous}.
Through the use of this correlation function, we demonstrate that the hop direction of the particles is encoded in the structure of the cage. Our measurements of the length scale associated with the new correlation function clearly reveals significant changes in the medium-range structural order of DH across $T_o$ and $T_c$, indicating that these crossover temperatures have an underlying thermodynamic origin and are not just temperatures that can only be extracted from fits to dynamic data.

\section{Results}

{\it System:} The system we investigate is the binary Lennard-Jones mixture from Ref.~\cite{KobAndersen1994}, slightly modified in order to prevent crystallization even at the lowest temperatures~\cite{schroder2020solid}. Previous studies determined $T_o$ and $T_c$ to be 0.8 and 0.435, respectively~\cite{KobAndersen1995, kob1995testing} and here we probe the equilibrium properties of the system down to $T=0.375$. The number of particles is 32400, and the box size is 30.0 in reduced units, which corresponds to a density of 1.2. Simulation details are provided in the Methods section, and basic structural and dynamic quantities are presented in Fig.~S1.

\begin{figure*}[!ht]
	\centering
	\includegraphics[width=1.0\textwidth]{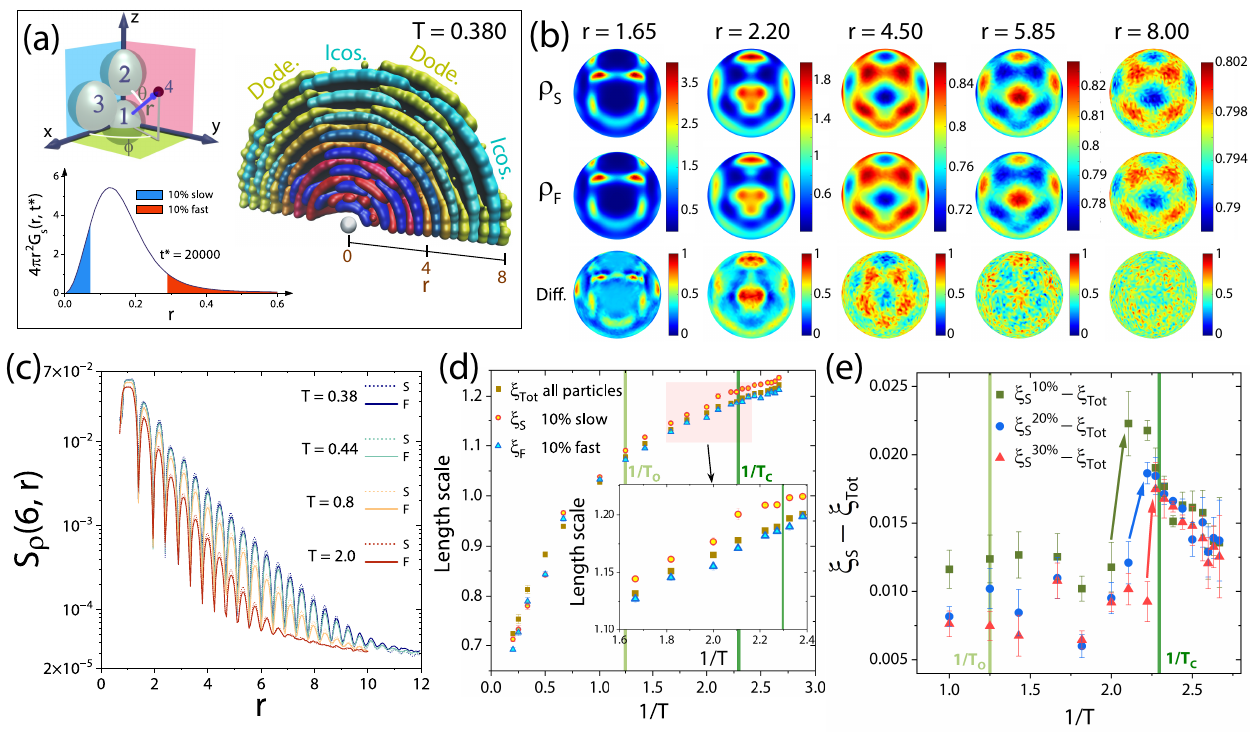} 
	\caption{{Connection between particle mobility and structural length scales.} 
(a) Schematic of the local coordinate system used to define the four-point correlation function. The right panel shows the three-dimensional density distribution of particles for $T = 0.38$ for $r$ values extending up to $8$. The shells shown correspond to local maxima of density. Shell thickness is $0.4$. Only the top 40\% of the density in each shell is shown. The bottom left panel shows the probability distribution of particle displacements, the self-part of the Van Hove function $G_s(r,t)$ at time $t^*$, the maximum of the non-Gaussian parameter. The shaded red (blue) region represents the top 10\% fastest (slowest) particles.
(b) Density maps of top 10\% slowest and top 10\% fastest particles (top and middle row, respectively) at different $r$ projected on the unit sphere for $T = 0.38$. The bottom row shows the difference field $\rho_D=\rho_S-\rho_F$.
(c) The angular power spectra of the density $S_\rho(6, r)$ of the top 10\%  fastest/slowest particles at different temperatures decay exponentially before reaching the noise threshold for large $r$.
(d) Length scales corresponding to the decay of $S_\rho(6, r)$ for top 10\% fastest/slowest and all particles ($\xi_{\rm Tot}$) at $t = 0$. Inset shows a zoom on this data near $T_{c}$. Error bars represent the standard error of the mean.
(e) Difference in structural length scales for the lowest 10\%, 20\%, and 30\% mobility cut-offs of slow particles and $\xi_{\rm Tot}$ for different temperatures. Arrows represent the sharp jump in this difference seen near $T_c$. Error bars show the standard error of the mean estimated using the jackknife method.
}
	\label{fig:1}
\end{figure*}

\noindent {\it 3D structure:} We probe the 3D structure of the liquid by constructing a local coordinate system (LCS) defined by a triplet of large particles that are mutual nearest neighbors (see Methods)~\cite{zhang2020revealing}. Within each triplet, we place particle \#1 at the origin, particle \#2 on the $z$-axis, and particle \#3 in the $x$-$z-$plane (Fig. \ref{fig:1}(a)). In this LCS, we calculate the probability of finding a fourth particle at spherical coordinates $(r,\theta,\phi)$. Averaging over all such triplets (different orientations of the LCS) improves statistical accuracy, and the resulting 3D density field from this four-point correlation function, $\rho_{\rm Tot}(r,\theta,\phi)$, is shown in Fig. \ref{fig:1}(a) for $T = 0.38$ (see Fig.~S2(a) for other temperatures). This field is highly anisotropic, comprising alternating layers of icosahedral and dodecahedral symmetries that are rotated with respect to each other, see Fig.~S2(a), in agreement with previous studies~\cite{zhang2020revealing, singh2023intermediate}. (We note that the order in the first nearest neighbor shell is not quite icosahedral because of chemical short-range order~\cite{coslovich2007understanding}. However, for larger distances this short range order is not relevant.) This order arises because particles in each subsequent shell occupy the hollows left by those in the preceding one, and the icosahedron and dodecahedron are geometric duals. Further, this structure gets more well-defined with supercooling, extending beyond $r=8$ for $T<T_c$, see Fig.~S2. 

Next, we extend this approach to examine the structure of DH. We identify slow (fast) particles as the 10\% of particles with the smallest (largest) displacements measured at time $t^*$ - defined by the location of the maximum in the non-Gaussian parameter $\alpha_2$ (see Supplementary Text~\cite{supple},~Eq..~S4), i.e., when the difference between slow ($S$) and fast ($F$) particles is largest (see Methods and Fig.~S1(e) and (f); Fig.~S1(e) shows that with decreasing $T$ the ratio $\tau_\alpha/t^*$ increases from a value around 1.0 to 70)~\cite{donati1998stringlike}. The density fields of these particles, $\rho_S$ and $\rho_F$ (Fig. \ref{fig:1}(b)), are qualitatively similar and resemble the total density field $\rho_{\rm Tot}$, Fig.~S2(b). However, the difference between these fields, $\rho_D = \rho_S - \rho_F$ (bottom row of Fig.~\ref{fig:1}(b)), is also highly anisotropic, confirming that the slow particles have a more ordered environment than the fast ones — a finding consistent with recent colloid experiments~\cite{singh2023intermediate}.  Importantly, this structural difference is not limited to small $r$ but extends to the intermediate range, showing that particle mobility correlates with structure far beyond the local cage. It is only beyond surprisingly large length scales, $r\gtrsim 8$ for $T = 0.38$, that $\rho_D$ vanishes.

To quantify this structural difference, we decompose $\rho_{\rm Tot}$, $\rho_S$, and $\rho_F$ into spherical harmonics and calculate the angular power spectrum, $S_{\rho}(l,r)$, for different order $l$, see Methods. Consistent with the observed icosahedral and dodecahedral symmetries in these density fields, the angular power is maximal for $l = 6$ for all distances~\cite{zhang2020revealing,singh2023intermediate}, Fig.~S2(c). Remarkably, the difference in $S_\rho(6,r)$ between $S$ and $F$ particles is modest at small distances and most pronounced for intermediate $r$ values, see Fig. \ref{fig:1}(c), highlighting that the mobility of a particle is determined by the structure on intermediate length scales.

The exponential $r$-dependence of $S_\rho(6,r)$ allows extraction of structural correlation lengths $\xi_{\rm Tot}$, $\xi_S$, and $\xi_F$, Fig.~\ref{fig:1}(d). (See Figs. S3 and S4 for different thresholds; Fig.~S5(e) shows that $\xi$ grows like $\log(1/T)$).)
Upon lowering $T$, all three length scales first grow rapidly, but this growth slows down in the supercooled regime, i.e., for $T\lesssim$ $T_o$. For $T\leq1.25$, $\xi_S$ is consistently larger than $\xi_F$, indicating that the slow regions are more ordered than the fast ones (Figs. \ref{fig:1}(d), S5(b), and (c)). Remarkably, while $\xi_{\rm Tot}$ and $\xi_F$ grow smoothly, $\xi_S$ shows a step-like increase near $T_c$ (see Inset to Fig. \ref{fig:1}(d)) - a clear structural signature of the dynamic crossover. Importantly, this step is only detectable in our $S_\rho$ analysis; correlation lengths associated with slow and fast particles from conventional observables like the pair correlation function $g(r)$, fail to reveal this jump (Fig. S5(a)). This failure explains why no previous study detected $T_c$ in structural quantities.

The step in $\xi_S$ manifests as a peak in the $T$-dependence of the difference $\xi_S-\xi_{\rm Tot}$ (Fig.~\ref{fig:1}(e)), which suggests that the slow particles form a network that undergoes a percolation-like transition near $T_c$. Two additional facts reinforce this inference: The peak position coincides with the temperature where the cluster size of DH is maximal (Fig.~S6(c)), and, second, the peak shifts towards $T_c$ on raising the mobility threshold to include more particles in the immobile set, Fig.~\ref{fig:1}(e). The rationale for this shift is that near $T_c$ the particles selected with a low threshold form a percolating structure that is structurally very organized and differs substantially from the average structure. If, however, the threshold is increased, the structure within the slow domains will, at the same $T$, differ less from the average one. It is only when the $T$ is lowered, that these slow regions will have formed a structure that is significantly better organized than the average one, i.e., the difference $\xi_S-\xi_{\rm Tot}$ will increase, although this increase will be inferior compared to the one observed with the smaller threshold, in agreement with the curves shown in Fig.~\ref{fig:1}(e). This interpretation of a percolating backbone composed of immobile particles is compatible with the double percolation scenario proposed recently~\cite{gao2025unified}. Such a backbone will endow the sample with rigidity while still permitting structural relaxation, consistent with experiments reporting liquid-like pockets even below $T_g$~\cite{cornet2023denser, chang2022liquid}. Note that the jump in $\xi_S$ seen in Fig.~\ref{fig:1}(e) is only around 2\%. However, from Fig.~\ref{fig:1}(d) one recognizes that at low $T$s an increase of the length scale $\xi_{\rm Tot}$ by 2\% corresponds to a decrease in temperature that is substantial, which in turn implies an increase of the relaxation time by a large factor. One thus concludes that this enhancement of the  correlation length can have a significant impact on the relaxation dynamics.
\\\\

\noindent {\it Anisotropy of the cage:} Figure~\ref{fig:1}(d) shows that the observable $S_\rho$ signals no structural change at $T_c$ for the fast particles. A possible reason is that the averaging of the density field over different orientations of the LCS washes out such a signature. Such an averaging is only appropriate when the environment around the fast particles is isotropic, as was assumed in Ref.~\cite{singh2023intermediate}. Since fast particles escape from their cages, the structural symmetry must be expected to be broken, and hence $S_F$ is not an appropriate quantity to probe their structure.

\begin{figure*}[!ht]
\centering
\includegraphics[width=0.8\textwidth]{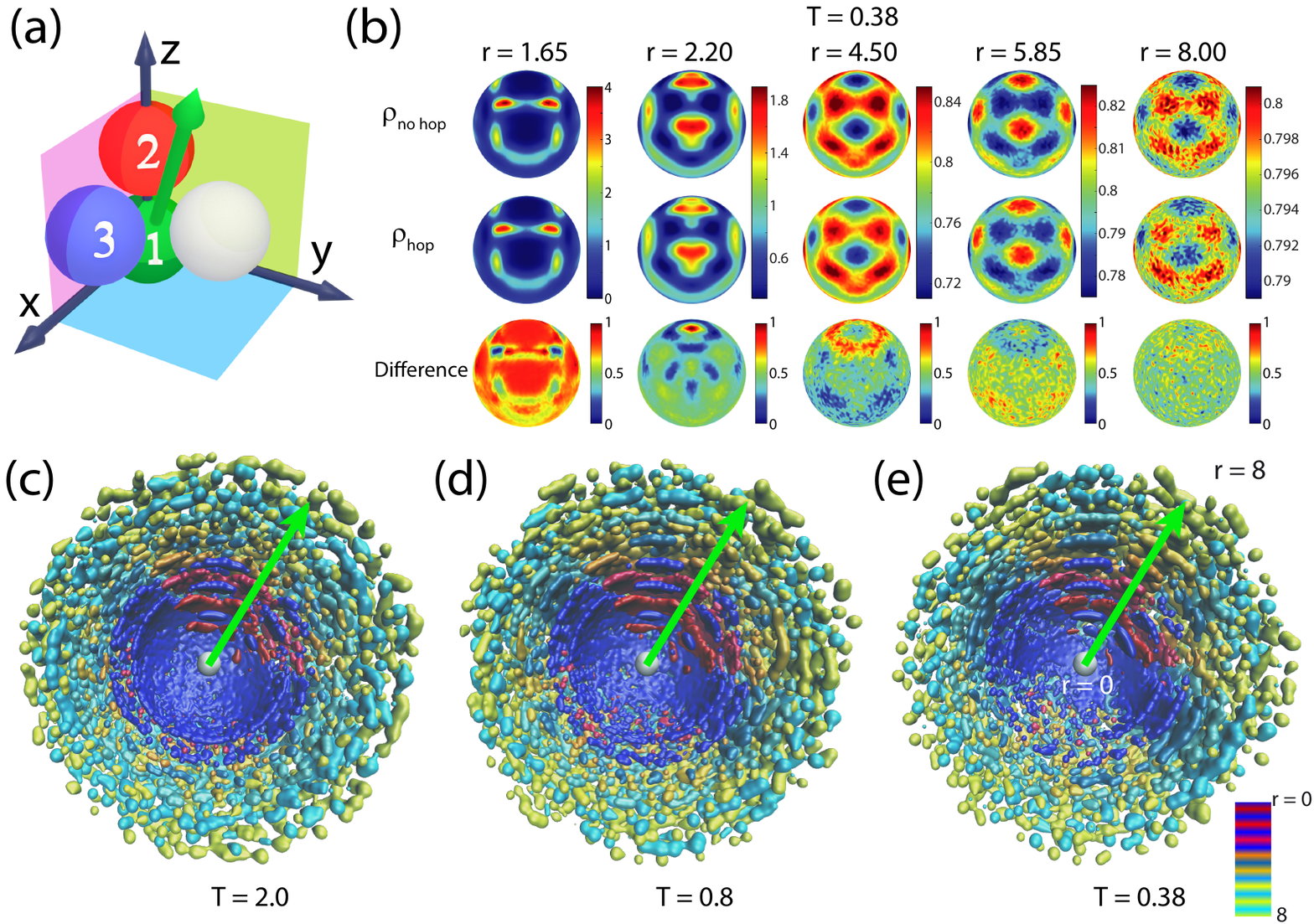}
\caption{Particle cage structure is anisotropic and extends to medium-range distances.
(a) The local coordinate system with the tagged particle (\#1, green) at the origin, the particle \#2 (red) is closest to the hopping direction (marked by the green arrow) and defines the $z$-axis.
(b) Normalized density distribution of particles in the local coordinate system. Top and middle rows show the density distribution around particles that did not hop (top 10\% slow) and around particles that hopped (top 10\% fast), respectively. The bottom row shows the difference $\rho_D^H$ in structure between the $S$ and $F$ particles.
(c-e) Three dimensional representation $\rho_D^H$ for a high temperature liquid ($T = 2.0$), at the onset temperature ($T = 0.8$), and at a temperature well below $T_c$ ($T = 0.38$). The vector represents the direction of hop and lies on the cutting plane of the displayed cross-section. For clarity, we show only the top $40\%$ of $\rho_D^H$ values, with the shell thickness increasing linearly with $r$.
}
\label{fig:2}
\end{figure*}

The relevance of this anisotropy can be probed by singling out the LCS that is best aligned with a particle's future hop direction. For a particle of interest (placed at the origin, \#1 in Fig.~\ref{fig:2}(a)), we identify its nearest neighbors at $t = 0$, and then orient the $z$-axis of the LCS along the neighbor (\#2) that is closest to the hop direction of \#1 at time $t^*$ (shown as green arrow). Placing particle \#3 - the nearest neighbor of both \#1 and \#2 - in the $x$-$z-$plane, fully determines the LCS. Since for times $t \leq t^*$, slow particles only rattle inside their cage, their displacement at $t^*$ points in a random direction, constituting a reference case. In this hop-aligned LCS, we calculate at $t = 0$ the instantaneous density fields of mobile ($\rho_{\text{hop}}$) and immobile ($\rho_{\text{nohop}}$) particles as above, i.e., before the particle leaves its cage. 
Note that since we are quantifying the instantaneous density fields, we are probing here a purely static quantity, namely the structural environment around these two classes of particles. We mention that the density field that is probed contains A and B particles. Qualitatively the same results are obtained if one considers only A particles, see Fig. S9.

\begin{figure*}[!ht]
\centering
\includegraphics[width=0.8\textwidth]{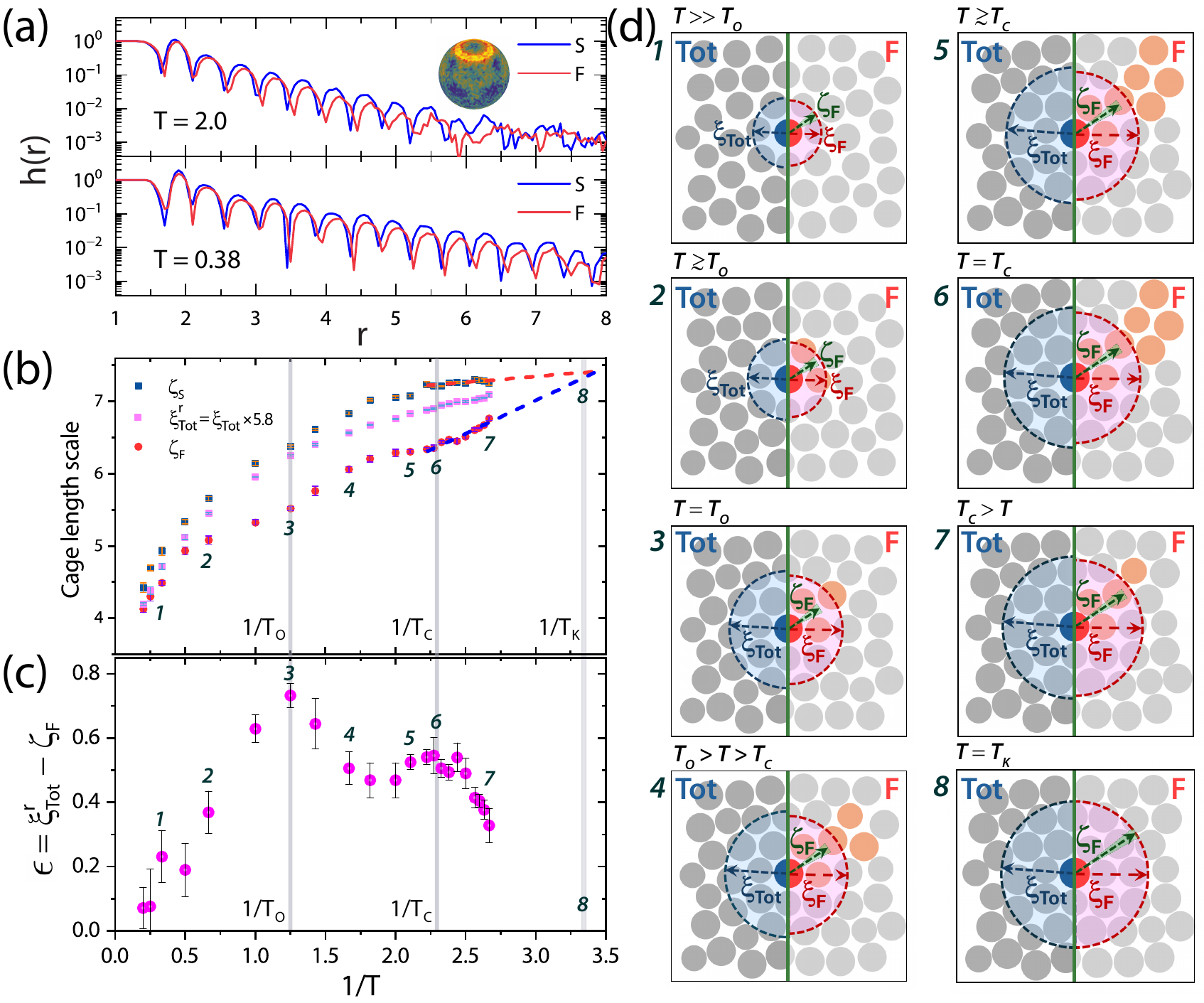}
\caption{Evolution of the asymmetric cage structure upon cooling.
{(a)} Directional correlation function, $h(\bf{r})$, for the slow (blue curves) and fast (red curves) particles at $T = 2.0$ and $T=0.38$, i.e., below $T_c$.
{(b)} $T-$dependence of length scales extracted from the slope of $h(\bf{r})$ for slow ($\zeta_S$) and fast ($\zeta_F$) particles shown by blue squares and red circles, respectively. The magenta squares represent $\xi_{\rm Tot}$ scaled by a factor of 5.8. 
The red and blue dotted lines are extrapolations that intersect near $T_K$. Error bars represent the standard error of the mean. 
{(c)} $T-$dependence of the cage asymmetry parameter $\epsilon = \xi^{r}_{\rm Tot} - \zeta_F$. Error bars show the standard error of the mean estimated using the jackknife method. 
{(d)} Schematic comparing the $T-$dependence of $\xi_{\rm Tot}$, $\xi_F$ and $\zeta_F$. To facilitate this comparison, the left (blue) half of the particle of interest depicts a typical particle in the system, and the right (red) half of the particle depicts a fast particle. The radii of the blue and red semicircle represent the structural length scales $\xi_{\rm Tot}$ and $\xi_F$, respectively. The green arrow indicates the cage escape direction, with length $\zeta_F$. Orange circles mark particles with excess disorder along the escape direction. 
}
\label{fig:3}
\end{figure*}

This approach reveals that the cage-breaking process is associated with a striking asymmetry of the structure: while $\rho_{\text{hop}}$ and $\rho_{\text{nohop}}$ retain the pronounced icosahedral - dodecahedral symmetries found in $\rho_S$ and $\rho_F$ (top and middle rows in Fig.~\ref{fig:2}(b)), the difference $\rho_D^H = \rho_{\text{nohop}} - \rho_{\text{hop}}$, does not (bottom row of Fig.~\ref{fig:2}(b)). Instead, $\rho_D^H$ shows 
a marked modification in particle density concentrated within a conical region centered around the $z$-axis, with the hop direction being the generator of this cone. The 3D representations of $\rho_D^H$ across several temperatures (Fig.~\ref{fig:2}(c-e)) bring this asymmetric cage structure to life. By orienting the $z$-axis and the hopping direction in the plane of the paper, these renderings expose a channel of weak structural order that becomes the future escape path of mobile particles out of their cages. (Note that for this representation we consider only LCS for which the hopping direction is in one fixed octant.) This escape channel grows in extent as the temperature decreases, and for $T<T_c$, stretches to distances $r>8$. Importantly, our analysis reveals that cage-breaking - the most elementary relaxation event in a supercooled liquid - is not a local phenomenon as originally thought, but is highly non-local, unlike crystalline systems wherein diffusion is a local, vacancy-mediated process~\cite{shewmon2016diffusion}. This insight rationalizes the difficulty in developing a robust theoretical description of relaxation dynamics in glass-forming systems at low temperatures, since previous analytical approaches have only considered relaxation processes that neglect the local anisotropy of the structure. Independently, this anisotropic and extended local disorder indicates that facilitated relaxation dynamics must be non-local and many-body, yet one-dimensional in character~\cite{keys2011excitations}.

We characterize the structure of the escape path via the directional correlation function, $h({\bf r}) = |g({\bf r})-1|$, which we average over all vectors $\bf r$ that are inside the conical region formed by the hopping direction mentioned above, i.e., have an opening angle between $23^\circ$ and $33^\circ$ with the $z-$axis. For comparison, we repeat this analysis for the $S$ particles with $\bf r$ pointing along the displacement vector of the particle at $t^*$, which is basically a random direction. Figure~\ref{fig:3}(a) and Fig.~S7 show that for all temperatures, $h({\bf r})$ is independent of particle mobility at small $r$, while with increasing $r$ the curve for the $S$ particles is above the one for the $F$ particles, confirming the enhanced ordering around the former. Moreover, $h({\bf r})$ decays faster for the $F$ particles, supporting the visual evidence presented in Figs.~\ref{fig:2}(c-e). The decay is exponential, allowing us to extract a structural length scale of the cage for the slow ($\zeta_S$) and fast ($\zeta_F$) particles. (In practice, this is done by requiring that the correlator has reached the threshold $10^{-2}$. Note that this definition of the length scales makes that the absolute value of $\zeta$ is about a factor of 6 larger than the one of $\xi$ presented in Fig.~1(d). However, since the correlator decays exponentially, this factor is basically constant and does not affect the following analysis. In the SI we show that the choice of the timescale used to select particle, i.e., $\tau^*$, is not crucial, Fig.~S8.) Importantly, the difference in $h({\bf r})$ for fast and slow particles vanishes along other directions, like in the equatorial plane, see Figs.~S7(c) and S7(d), demonstrating that the structural difference between these two particle species is primarily along the cage escape direction, in agreement with Fig.~\ref{fig:2}(b).

On cooling, $\zeta_S$ (Fig.~\ref{fig:3}(b)) mirrors the $T-$dependence of $\xi_S$ (Fig.~\ref{fig:1}(d)), including the step at $T_c$, because $S$ particles do not break cages for $t \leq t^*$, and hence directional and orientationally-averaged correlations are basically equivalent. In contrast, the $T$-dependence of $\zeta_F$ is strikingly different -- its growth is interrupted by two pronounced slowdowns, first near $T_o$ and then near $T_c$ (Fig. \ref{fig:3}(b)), before resuming below $T_c$ its growth in a more rapid way. Remarkably, extrapolation of $\zeta_S$ and $\zeta_M$ to lower temperatures (dashed lines) indicates that they meet near the Kauzmann temperature, $T_K\approx 0.30$~\cite{sciortino1999inherent}, the putative temperature where the distinction between slow and fast particles no longer exists since relaxation dynamics ceases completely.

Close to the crossover temperatures, $\zeta_F$ is almost $20\%$ smaller than $\zeta_S$, a much larger difference than the few percent observed between $\xi_F$ and $\xi_S$, indicating that directional structural measures are indeed better suited to reveal relevant length scales, see also Ref.~\cite{hu2026anisotropic}. The $T-$dependence of $\zeta_F$ can be appreciated well by comparing it with the mean structural length $\xi_{\rm Tot}^r=5.8\xi_{\rm Tot}$, where the rescaling is done to yield a comparable magnitude to $\zeta_F$, see Fig.~\ref{fig:3}(b). The temperature evolution of the cage asymmetry parameter $\epsilon=\xi_{\rm Tot}^r-\zeta_F$ is presented in Fig.~\ref{fig:3}(c) and shows that at $T_o$ and $T_c$ this difference has pronounced maxima, i.e., $\zeta_F$ is small with respect to the system average. This graph clearly demonstrates that it is possible to detect the two crossover temperatures from purely structural quantities, i.e., these temperatures are not just related to dynamic quantities. This important insight can be rationalized as follows. At the highest $T$s the structure around the $F$ and $S$ particles differs only little from the average structure in the liquid, since it is essentially isotropic (Fig. \ref{fig:3}(d)1), and hence $\epsilon$ is small. Upon cooling, the system becomes more ordered, i.e., cage formation begins. Due to structural fluctuations, some particles are in less ordered environments, thus having directions that serve as preferential pathways along which they can escape sooner than the average particle (Fig. \ref{fig:3}(d)2). These particles form the fast population. On approaching the onset temperature, $\xi_{\rm Tot}$ grows faster than $\zeta_F$ ($\epsilon$ increases) (Fig. \ref{fig:3}(d)3), implying that while on overall the cage becomes more ordered (rigid), the disorder along the weak directions for the $F$ particles changes only a little. Below the onset temperature, $\epsilon$ decreases, i.e., $\zeta_F$ grows faster
than $\xi_{\rm Tot}$ (Figs.~\ref{fig:3}(b) and \ref{fig:3}(d)4). This indicates that the environment around the fast particle will, upon cooling, mainly improve its order by restructuring the quasi-1D leaky channels. Importantly, because $\zeta_F<\xi_{\rm Tot}$, these leaky paths are always less ordered compared to the system average and hence permit a higher particle mobility. For temperatures below $T_o$, the $F$ particles exploit this enhanced structural disorder and escape their cages along these unstable directions. Previous studies found that in this $T-$regime the relaxation mechanism takes the form of string-like \textit{cooperative} particle motion~\cite{donati1998stringlike, weeks2000three, stevenson2006shapes, hima2015direct}. The asymmetric structure of the cage (Fig. \ref{fig:2}(d)) thus provides a natural structural basis for this type of collective dynamics. A maximum in $\epsilon$ at $T_o$ thus signals a change in transport mechanism for the $F$ particles: From one governed by strong structural disorder above $T_o$ to that associated with increasingly cooperative relaxation dynamics below $T_o$. This interpretation of a change in the mechanism is consistent with recent experiments on colloidal systems~\cite{li2020anatomy}, where local perturbations relaxed through the formation of complex dynamical structures, whose size peaks at $T_o$. 

On cooling towards $T_c$, $\xi_{\rm Tot}$ continues to grow slowly, while $\zeta_F$ levels off ($\epsilon$ increases) (Figs. \ref{fig:3}(b) and \ref{fig:3}(d)5/6). This plateau in $\zeta_F$ implies that the string-like excitations stop growing and no longer constitute an efficient relaxation channel. This conclusion aligns with previous studies that found the nature of cooperativity undergoing a morphological change from string-like to compact dynamical structures near $T_c$~\cite{stevenson2006shapes, kob2012non, hima2015direct, ortlieb2023probing}. Below $T_c$, $\zeta_F$ steadily increases to approach the value of the correlation length in the system, and consequently, $\epsilon$ decreases again. As the escape path becomes as ordered as its surroundings, the environment around the $F$ particles becomes more symmetric again (Fig. \ref{fig:3}(d)7), and relaxation proceeds predominantly via a simpler hopping mechanism~\cite{rusciano2025low, sastry1998signatures}. Our findings suggest that structural anisotropy, absent in the high $T$ liquid, is also absent at $T=T_K$ (Fig. \ref{fig:3}(d)8), revealing a striking symmetry that links the liquid and the `\textit{ideal}' glass. 

\section{Discussion}

The conceptual advance presented here, i.e., the introduction of a local coordinate system that takes into account the future escape direction of the particle from its cage, allows determining crossover temperatures solely from structural observables. This demonstrates that $T_o$ and $T_c$ are not merely fit parameters to transport quantities or theory-dependent constructions, but are instead intrinsic to the system and hold special significance in the relaxation dynamics of glass-forming liquids. Unlike structural correlation lengths that grow smoothly as the glass-transition temperature $T_g$ is approached, those associated with the slow particles ($\xi_S$) and cage anisotropy ($\zeta_F$) do not since they track changes in the transport mechanism at $T_o$ and $T_c$. Such a strong structure–dynamics link shows that the critical temperature of MCT should not be considered merely a fit parameter for describing dynamic quantities, but rather that it has deep implications for the structure. 
While earlier studies have discussed various system-specific local structural descriptors that show a noticeable change close to $T_o$, the correlation length presented here is agnostic, i.e., can be used for any glass-forming system. In addition, and most important, this correlation length is able to predict the location of $T_c$, in contrast to any structural descriptor considered in earlier studies. Although this discovery gives strong support for the soundness of MCT, it does, however, not exclude the presence of other types of relaxation mechanisms that may be at work in parallel or at lower/higher temperatures.

The temperature evolution of $\zeta_F$ and $\xi_S$ identified here, allows to unify and quantify  several previously disconnected observations: The emergence of cooperative string-like motion below $T_o$, the change in morphology of cooperatively rearranging regions near $T_c$, and the development of connected immobile particle domains consistent with percolation-like scenarios. This ability to rationalize this variety of key features of glass-forming liquids shows that locally oriented coordinate systems are a highly effective tool to obtain insight into the relation between structure and dynamics of glass-formers, an important step forward in our understanding of the slow dynamics of glass-forming systems. The approach outlined here can be readily implemented in experimental systems where trajectories are directly measurable, such as colloidal suspensions. Probing the crossover physics that our simulations have uncovered in these systems, however, requires new strategies for achieving precise control of particle number density, which at present is challenging~\cite{poon2012measuring}. Extension to other glass formers, including network and molecular liquids, will be essential to establish the generality of the structural signatures reported here and to determine which features are model-specific. Finally, we note that for molecular glass-formers one can use directly the (non-spherical) molecule to define a local coordinate system and then probe how the relaxation dynamics of the surrounding molecules depends on the relative position with respect to the central molecule.
The approach introduced here is thus a method that can be applied in a large class of glass-formers and hence will permit to advance our understanding of these complex systems.

\section{Methods}

\subsection*{System and simulation details} 
We simulate a binary (80:20) mixture of Lennard-Jones particles which has been studied extensively in previous works~\cite{kob1995testing}. To avoid crystallization at the lowest temperatures we use the modification of this model proposed in Ref.~\cite{schroder2020solid}.
The interaction potential between particles of type $\alpha$ and $\beta$ $\in$  $\{A, B\}$ is given by \\[-0.75mm]
\begin{equation}
V_{\alpha\beta}(r) = 4\varepsilon_{\alpha\beta}[(\sigma_{\alpha\beta}/r)^{12} - (\sigma_{\alpha\beta}/r)^{6}] \quad,
\end{equation}

\noindent
where $\sigma_{AA} = 1.0$, $\varepsilon_{AA} = 1.0$, $\sigma_{AB} = 0.8$, $\varepsilon_{AB} = 1.5$, $\sigma_{BB} = 0.88$ and $\varepsilon_{BB} = 0.5$. $\sigma_{AA}$ and $\varepsilon_{AA}$ are used as units of length and energy, respectively. The mass of the particles is set to $m = 1.0$, time is measured in units of $(m \sigma_{AA}^2/\epsilon_{AA})^{1/2}$, setting the Boltzmann constant $k_B = 1$. The AA and BB forces are shifted and truncated at $1.5\sigma_{AA}$ and the ones for AB at $2.5\sigma_{AB}$. The static and dynamic properties of this system are very similar to the model from Ref.~\cite{kob1995testing}, with no change in the crossover temperatures of interest ($T_{o} \approx 0.8$ and $T_c \approx 0.435$)~\cite{KobAndersen1994}.

We simulate 22 temperatures ranging from $T = 5.0$ to $T = 0.375$ at constant density $\rho=1.200$. Since the system is not prone to crystallization, the simulation is carried out for periods extending to $2\cdot10^9$ time steps ($h=0.005$) for $T\le 0.39$, well beyond the $\alpha$-relaxation time. To improve statistics, four independent simulations are carried out at every $T$. The total amount of single CPU time needed to obtain these trajectories is about 55 years.

Standard quantities characterizing the static and dynamic properties of the system are presented in Fig.~S1.

\subsection*{Decomposition of density into spherical harmonics}
The density distribution of particles in spherical coordinates, $\rho(\theta, \phi, r)$ determined via the local coordinate system (defined from three mutual neighbor particles of type A) can be decomposed into spherical harmonics $Y_l^m$:

\begin{equation}
\rho(\theta, \phi, r) = \sum_{l = 0}^{\infty} \sum_{m = -l}^{l}\rho_l^m(r) Y_l^m(\theta, \phi) \quad .
\end{equation}

The expansion coefficients $\rho_l^m$ are given by

\begin{equation}
\rho_l^m(r) = \int_{0}^{2\pi} d\phi \int_{0}^{\pi} \sin(\theta)\ \rho(\theta, \phi, r)\ Y_l^{m*}(\theta, \phi) \quad ,
\end{equation}

\noindent
where $Y_l^{m*}$ is the complex conjugate of the spherical harmonics of degree $l$ and order $m$. 

The angular power spectrum obtained from the expansion coefficients is then calculated from 
\begin{equation}
S_\rho(l, r) = \left[\frac{1}{2l+1}\sum_{m = -l}^{l}\left|\rho_l^m(r)\right|^2\right]^{1/2} \quad .
\end{equation}

For the numerical evaluation of $S_\rho(l,r)$ we have used for each run~100 configurations, spaced equidistant in time, i.e., a total of 400 configurations. While the particles used to define the local coordinate system are all of type A, the fourth particle making up the density field $\rho(\theta,\phi,r)$ can be of type A or B. 

\section*{Acknowledgments}
\paragraph*{Funding:}
 K.M. thanks the Jawaharlal Nehru Centre for Advanced Scientific Research (JNCASR), Bangalore, India, for a research fellowship. R.G. thanks JNCASR for financial support.
\paragraph*{Author contributions:}
K.M., R.G., and W.K. designed the research. W.K. did the simulations; K.M., R.G., and W.K. did the analysis of the data, interpreted the results, and wrote the manuscript.

\paragraph*{Competing interests:}
The authors declare that they have no competing interests.
\paragraph*{Data, code and materials availability:} 
All data needed to evaluate the conclusions in the paper are present in the paper and/or the Supplementary Materials.

\clearpage

\bibliography{mro}

\title{Supplementary material for \break ``Anisotropic medium-range order uncovers dynamic crossovers \break in glass-forming liquids"}

\author{Kamlesh Mishra}
\affiliation{Chemistry and Physics of Materials Unit, Jawaharlal Nehru Centre for Advanced Scientific Research, Bangalore 560064, India
}
\author{Rajesh Ganapathy}%
\affiliation{Chemistry and Physics of Materials Unit, Jawaharlal Nehru Centre for Advanced Scientific Research, Bangalore 560064, India
}
\affiliation{International Centre for Materials Science, Jawaharlal Nehru Centre for Advanced Scientific Research, Bangalore 560064, India}
\affiliation{School of Advanced Materials (SAMat), Jawaharlal Nehru Centre for Advanced Scientific Research, Bangalore 560064, India}
\author{Walter Kob}
\email{walter.kob@umontpellier.fr}
\affiliation{Department of Physics, University of Montpellier, CNRS, Montpellier F-34095, France}

\date{\today}



\newpage

\renewcommand{\thefigure}{S\arabic{figure}}
\setcounter{figure}{0}

\renewcommand{\theequation}{S\arabic{equation}}
\setcounter{equation}{0}

{\bf \Large Supplementary Information}

\section*{Temperature dependence of static and dynamic quantities}
\begin{figure}
\centering
\includegraphics[width=0.8\textwidth]{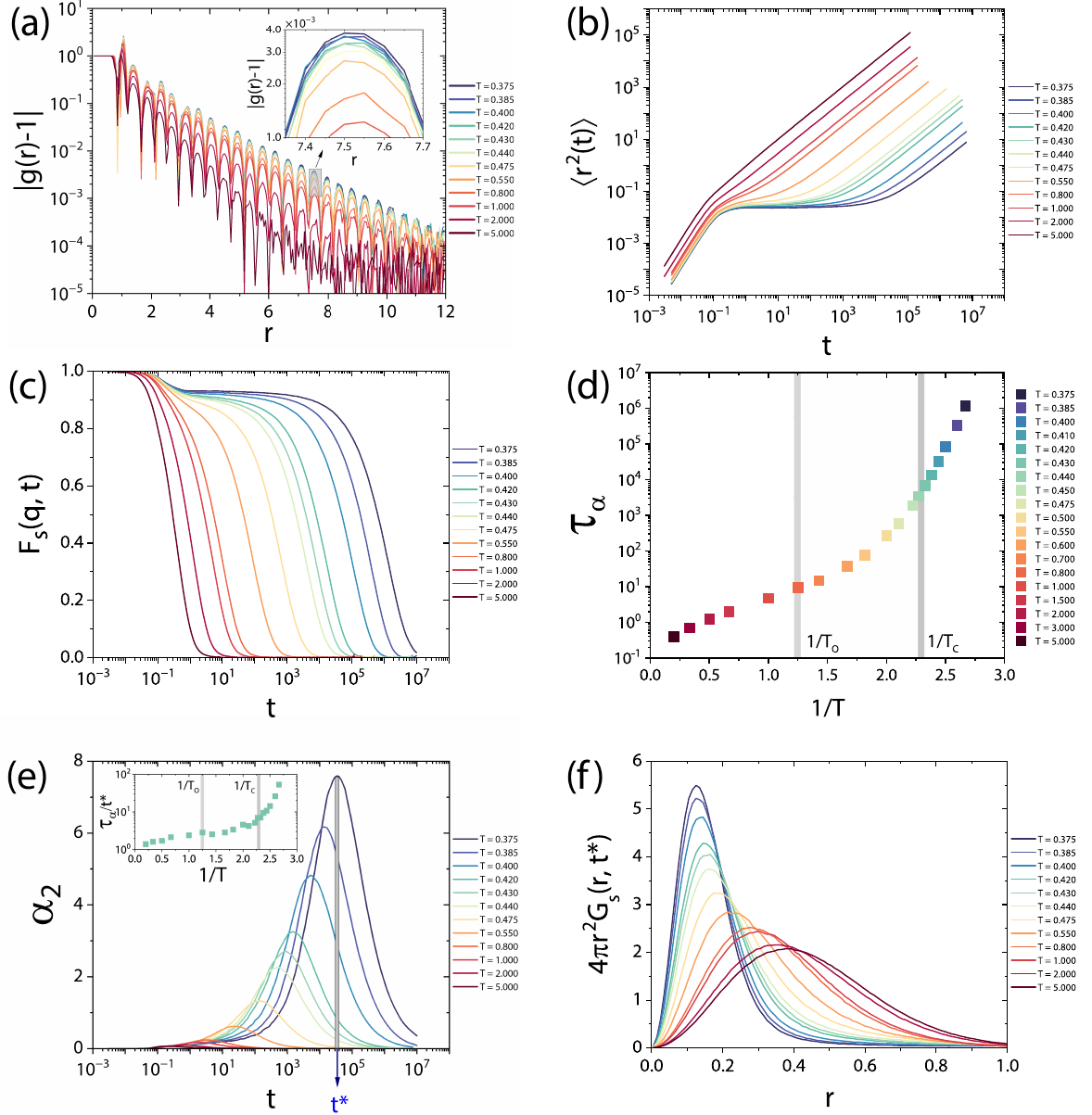}
\caption{Temperature dependence of static and dynamic quantities. (a) Radial distribution function, $\lvert g(r)-1\rvert$ for different temperatures. The inset shows a magnified view at an intermediate distance, $r\approx 7.5$ where the growth of peaks on cooling cannot be discerned. (b) The mean-squared displacement. (c) The self-intermediate scattering function, $F_s(q, t)$ for the wave-vector corresponding to the peak in the static structure factor. (d) Relaxation time, $\tau_\alpha$, corresponding to the time when $F_s(q,t)$ decays to $1/e$. The gray vertical lines indicate the crossover temperatures, $T_o$ and $T_c$. (e) The non-Gaussian parameter $\alpha_2$. The gray line shows the location of the peak ($t^*$) for the lowest temperature ($T = 0.375$). The inset shows the temperature dependence of the ratio of $\tau_\alpha$ and $t^*$. (f) The self-part of the Van Hove correlation function for $t= t^*$ for each temperature.}
\label{SI_fig:1}
\end{figure}

\noindent For all static and dynamic observables, we consider only type A particles, since they constitute the majority of the system. First, in Fig.~\ref{SI_fig:1}(a) we quantify structural order using the pair-correlation function, $g(r)$, given by~\cite{binder2011glassy}, 

\begin{equation}
    g(r) = \frac{V}{4 \pi r^2 N^2} \sum_{i=1}^{N} \sum_{j \neq i}^{N} \langle \delta(r - |r_i - r_j| ) \rangle \quad,
\end{equation}

\noindent
where $N$ is the total number of particles, and $V$ is the volume. In the inset to Fig. \ref{SI_fig:1}(a), we show the peak profile of $|g(r)-1|$ for an intermediate value of $r$. While the amplitude of the peak grows with cooling, for $T<T_{c}$, the insufficient signal-to-noise ratio makes it difficult to discern a clear trend.

We quantify the dynamics via the mean-squared displacement (MSD), 

\begin{equation}
\langle r^2(t)\rangle =  \frac{1}{N}\sum_{j = 1}^{N} \langle |\mathbf{r}_j(t) - \mathbf{r}_j(0)|^2 \rangle \quad,
\end{equation}

\noindent
presented in Fig.~\ref{SI_fig:1}(b), and the self-intermediate scattering function (SISF)

\begin{equation}
F_s(q, t) = \frac{1}{N}\sum_{j = 1}^{N} \langle e^{i\mathbf{q}\cdot(\mathbf{r}_j(t) - \mathbf{r}_j(0))} \rangle \quad,
\end{equation}

\noindent
shown in Fig.~\ref{SI_fig:1}(c)~\cite{KobAndersen1995, kob1995testing}.  Particle caging is evident as a plateau at intermediate times in both the MSD and SISF. The temperature dependence of the structural relaxation time, $\tau_\alpha$, defined as the time at which $F_s(q, t)$ has decayed to $1/e$, is shown in Fig.~\ref{SI_fig:1}(d). Particle displacements deviate most from Gaussian statistics as they break out of their cages and we extract the cage-breaking time $t^*$ as the time at which the non-Gaussian parameter, 
\begin{equation}
\alpha_2(t) = \frac{3 \langle r^4(t) \rangle}{5 \langle r^2(t) \rangle^2} - 1 
\end{equation}
is a maximum, shown in Fig. \ref{SI_fig:1}(e). The probability that a particle has moved a distance $r$ over $t^*$ is given by the self part of the Van Hove correlation function, 

\begin{equation}
G_s(r, t^*) = \frac{1}{N}\sum_{j = 1}^{N} \langle \delta (\mathbf{r}-|\mathbf{r}_j(t^*)-\mathbf{r}_j(0)|) \rangle
\end{equation} 
(Fig.~\ref{SI_fig:1}(f)) ~\cite{KobAndersen1995}. The tails of $G_s(r, t^*)$ correspond to particles that have undergone the smallest and largest displacements over $t^*$. For appropriately chosen cut-offs, these particles form the slow and fast particle species as shown in Fig. 1(a) of the main text.

\section*{Probing the structure of BLJM system using the four-point correlation function}

\begin{figure}
\centering
\includegraphics[width=0.8\textwidth]{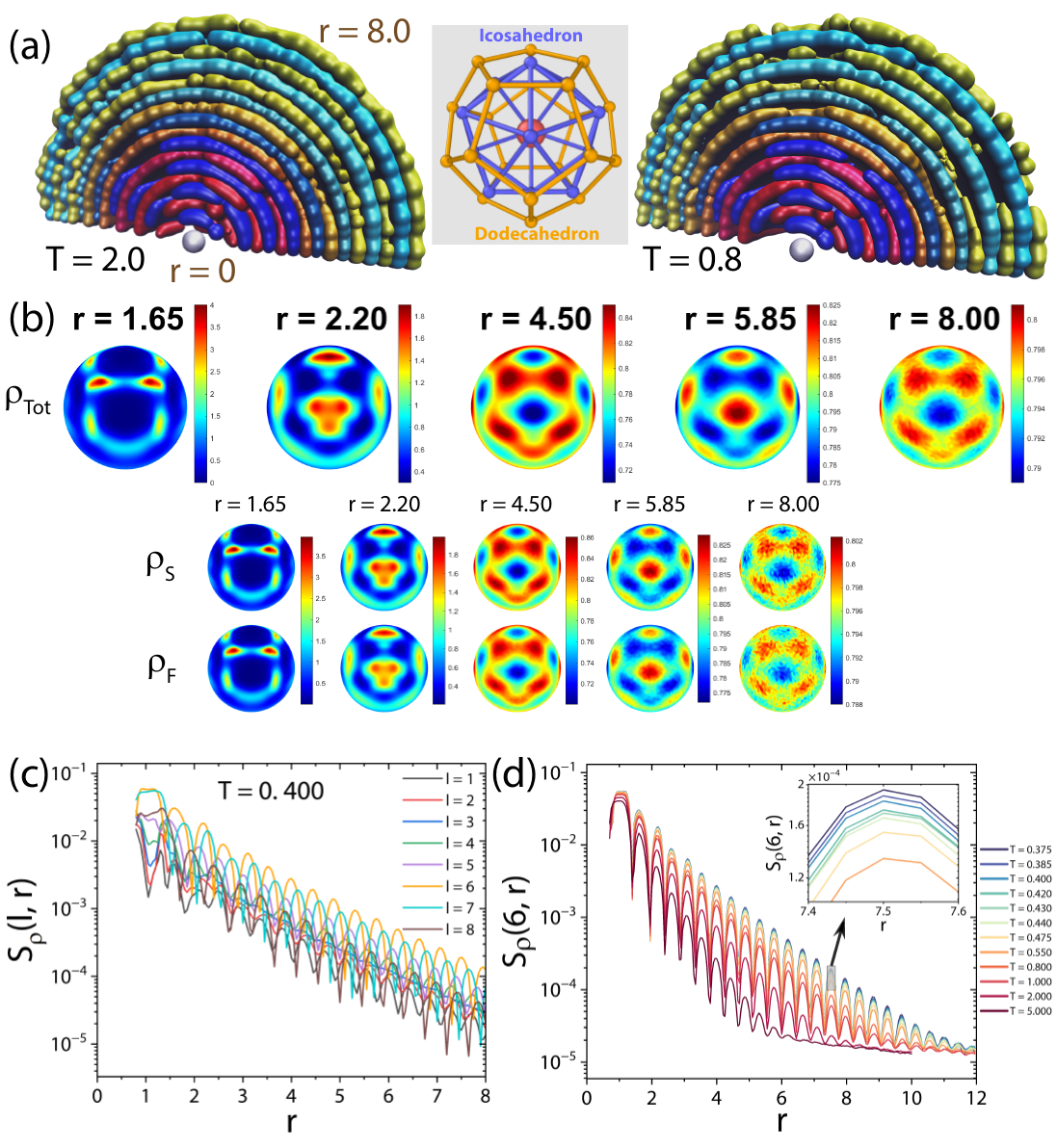}
\caption{Density distribution and angular power spectrum obtained using four-point correlation. (a) Three-dimensional density distributions for $T = 2$ and $T = 0.8$. Schematic shows the relative orientation of the dodecahedral and icosahedral density distribution. The dodecahedral and icosahedral symmetries seen in the density fields correspond to the location of the maxima and minima of $g(r)$. (b) Top row shows the density distribution averaged over all particles, $\rho_{Tot}$ for $T = 0.38$  for different $r$ values projected on unit sphere. Bottom two rows ($\rho_S$ and $\rho_F$) are reproduced from Fig. 1(b) of the main manuscript facilitating easy comparison with $\rho_{Tot}$. (c) Angular power spectrum, $S_\rho(l, r)$ corresponding to different $l$ values for a representative temperature $T = 0.4$. $l = 6$ has the strongest signal. (d) $S_\rho(6, r)$ for all temperatures. The inset shows zoomed in $S_\rho(6, r)$ near $r\approx 7.5$ showing distinct peaks for all temperatures.}
\label{SI_fig:2}
\end{figure}

\noindent In Fig.~\ref{SI_fig:2}(b), we 
show the three-dimensional density distribution, $\rho_{\rm Tot}(\theta,\phi,r)$, averaged over all particles, at  
$T = 0.38$ for different $r$ values projected on the unit sphere.
This distribution shows alternating layers of icosahedral and dodecahedral symmetries, in agreement with earlier results~\cite{zhang2020revealing}. The middle and bottom rows of the panel show the density field for the top $10\%$ slow ($\rho_S$) and fast ($\rho_F$) particles (reproduced from Fig. 1(b) of the main text). One recognizes that the structure around the $S$ and $F$ particles is very similar to the average structure.

Similar to Fig. 1(a) of the main text, we show $\rho_{\rm Tot}$ for two additional temperatures of $T = 2.0$ and $T = 0.8$ in Fig.~\ref{SI_fig:2}(a) and one clearly sees that the structural order increases with decreasing temperature. The decomposition of this density into spherical harmonics yields the angular power spectrum, $S_\rho(l, r)$. Figure~\ref{SI_fig:2}(c) demonstrates that for our system, $S_\rho(l, r)$ has the strongest signal for $l = 6$ at all $r$ values, consistent with the underlying icosahedral symmetry. We therefore focus in the main text on this $l$ value for all subsequent analyses. 

The decay of $S_\rho(6, r)$ becomes increasingly slow
on cooling, Fig.~\ref{SI_fig:2}(d). Unlike $g(r)$, Fig. \ref{SI_fig:1}(a), $S_6(l,r)$ has a significantly better signal-to-noise ratio (SNR). This is readily observed by noting that for the $r$ values where the $g(r)$ signal has reached its noise floor, the $S_\rho(6, r)$ still shows clearly discernible oscillations. The superior SNR is also apparent in the peak profile at intermediate $r$ values (inset to Fig.~\ref{SI_fig:2}(d), where the peak height shows a clearly discernible increase on cooling, even down to the lowest $T$. 

\section*{Probing the structure of dynamical heterogeneity using four-point correlation function}
\begin{figure}
\centering
\includegraphics[width=0.7\textwidth]{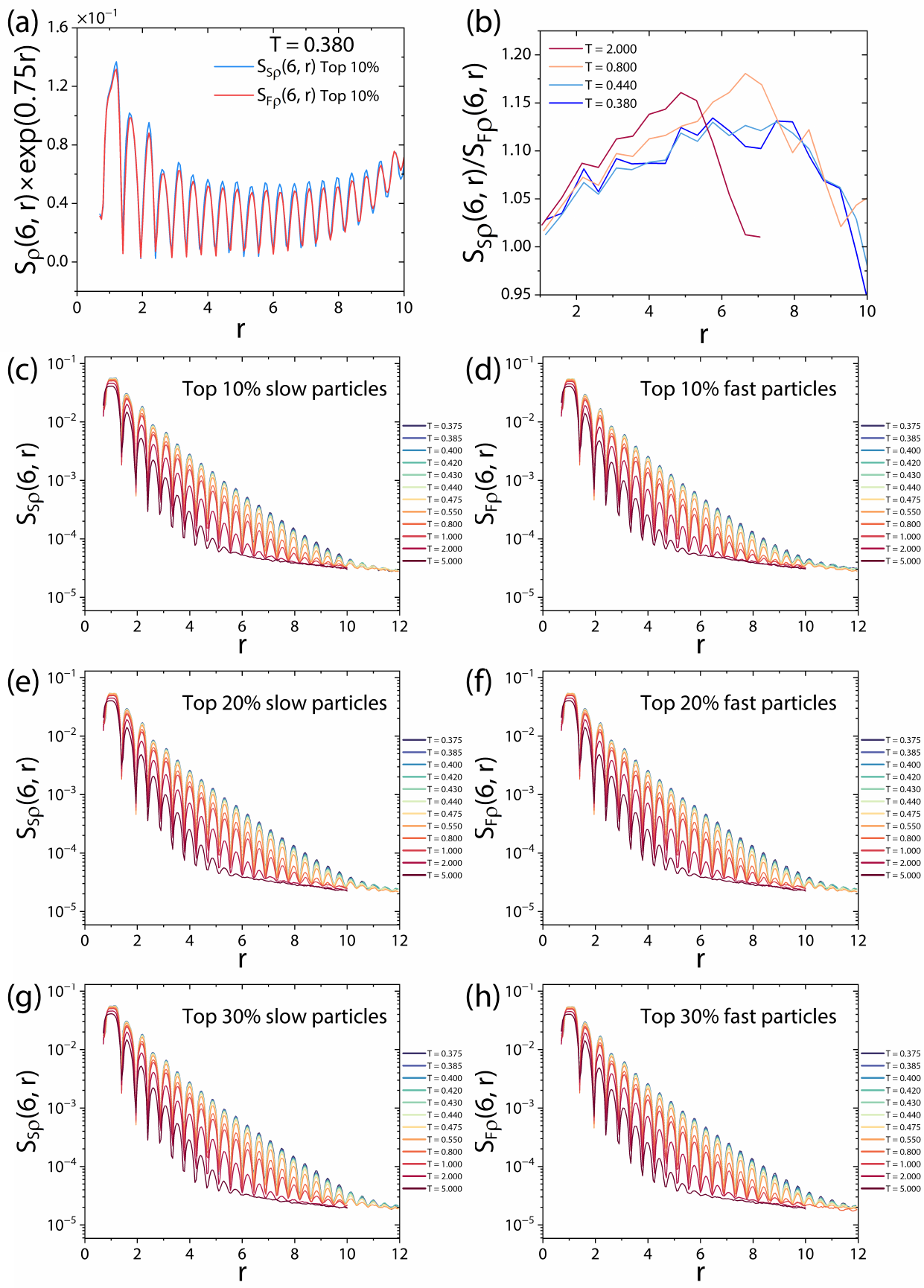}
\caption{Structure of dynamical heterogeneity. (a) Angular power spectrum for  slow ($S_{S\rho}(6, r)$) and fast ($S_{F\rho}(6, r)$) particles at $T = 0.38$ multiplied by an exponential $e^{0.75r}$. 
(b) Ratio of the peaks of $S_{S\rho}(6, r)$ and $S_{F\rho}(6, r)$ showing a maximum at intermediate $r$ value. (c-h) Temperature dependence of $S_{S\rho}(6, r)$ and $S_{F\rho}(6, r)$ for mobility cut-offs of 10\%, 20\% and 30\%. One sees that qualitatively the difference between the data for the $S$ and $F$ particles does not depend significantly on this cut-off. \\
}
\label{SI_fig:3}
\end{figure}

\noindent 
We select slow and fast particles using different cut-offs of the Van Hove function (Fig.~\ref{SI_fig:1}(f)). In Fig.~\ref{SI_fig:3}(a), we present the angular power spectrum for slow ($S_{S\rho}(6, r)$) and fast ($S_{F\rho}(6, r)$) particles multiplied by the exponential $e^{0.75r}$ for $T = 0.38$. This multiplication results in the height of the peaks being almost constant for $2 < r < 8$, confirming the exponential nature of the decay of $S_{\rho}(6, r)$. Importantly, the signal for the slow particles is above the fast ones, indicating the more ordered environment around the former. The difference in signal between $S$ and $F$ particles is most pronounced at the intermediate distances, consistent with the results from colloidal experiments~\cite{singh2023intermediate}. This behavior is better seen by plotting the ratio of the peaks of $S_{S\rho}(6, r)$ and $S_{F\rho}(6, r)$, which displays a maximum at intermediate $r$ values (Fig.~\ref{SI_fig:3}(b)). The location of this maximum shifts to larger values of $r$ as the system is cooled, again consistent with colloid experiments~\cite{singh2023intermediate}. 

\begin{figure}
\centering
\includegraphics[width=0.7\textwidth]{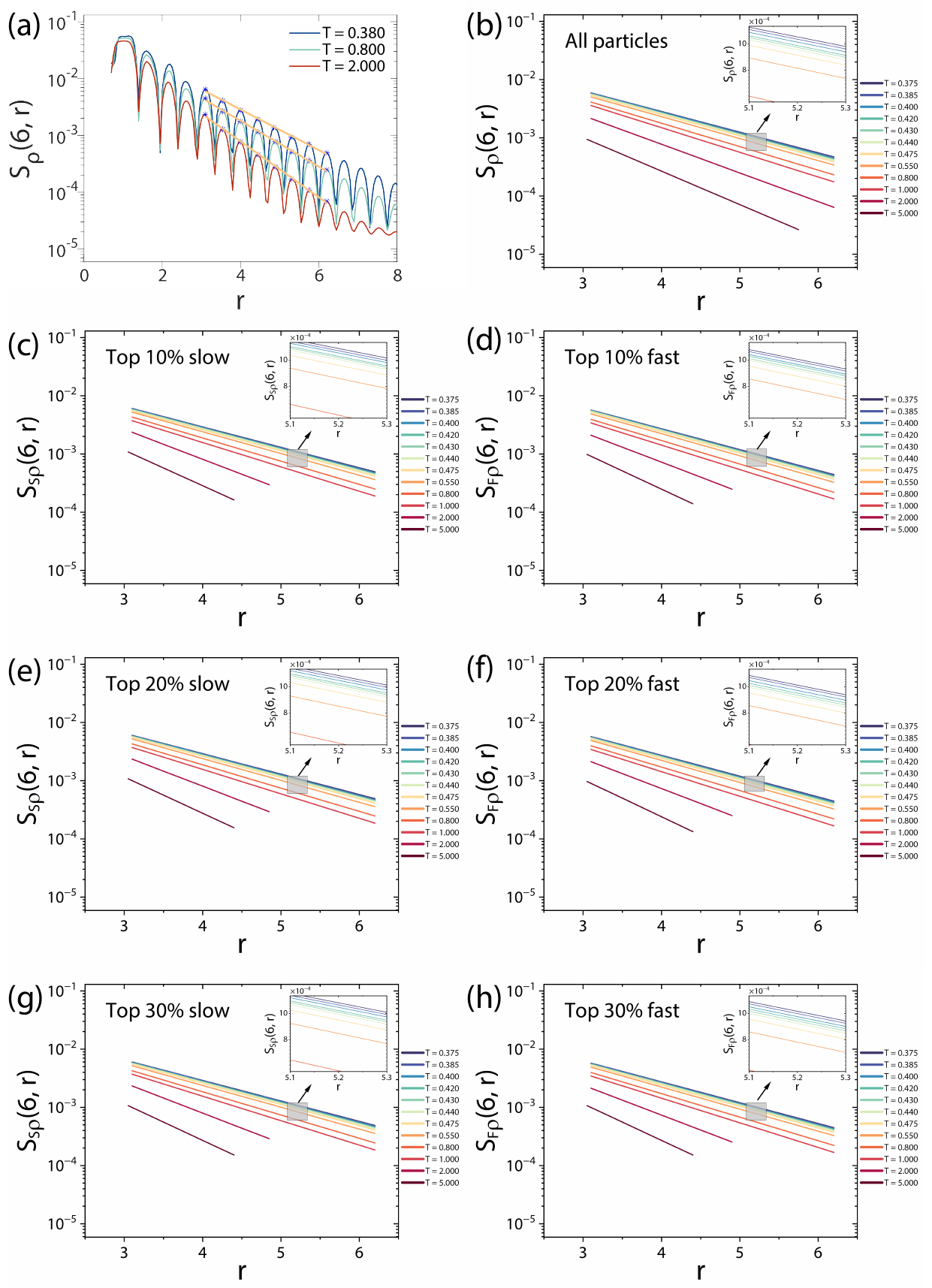}
\caption{Exponential fits to $S_\rho(6, r)$. Exponential fits of the form $S_\rho(6, r, T)\propto \exp[-r/\xi(T)]$ to the peaks of the angular power spectra shown in Fig.~\ref{SI_fig:2}(d) and Fig.~\ref{SI_fig:3}(c-h) at intermediate distances. The insets show magnified view of the smooth growth in slopes at low temperatures. \\
} 
\label{SI_fig:4}
\end{figure}

The temperature dependence of $S_{S\rho}(6, r)$ and $S_{F\rho}(6, r)$ is shown in Fig.~\ref{SI_fig:3}(c-h) for three different mobility cut-offs: 10\%, 20\% and 30\%. At all temperatures, height of the peaks
of $S_\rho(6, r)$,  $S_{S\rho}(6, r)$, and $S_{F\rho}(6, r)$ decay exponentially 
allowing to extract the corresponding structural length scales, $\xi_{Tot}$, $\xi_S$, and $\xi_F$ by fitting an exponential of the form $S_\rho(6, r, T)\propto \exp(-r/\xi(T))$ as shown in Fig.~\ref{SI_fig:4}.

\section*{Length scales associated with dynamical heterogeneity}
\begin{figure}
\centering
\includegraphics[width=0.8\textwidth]{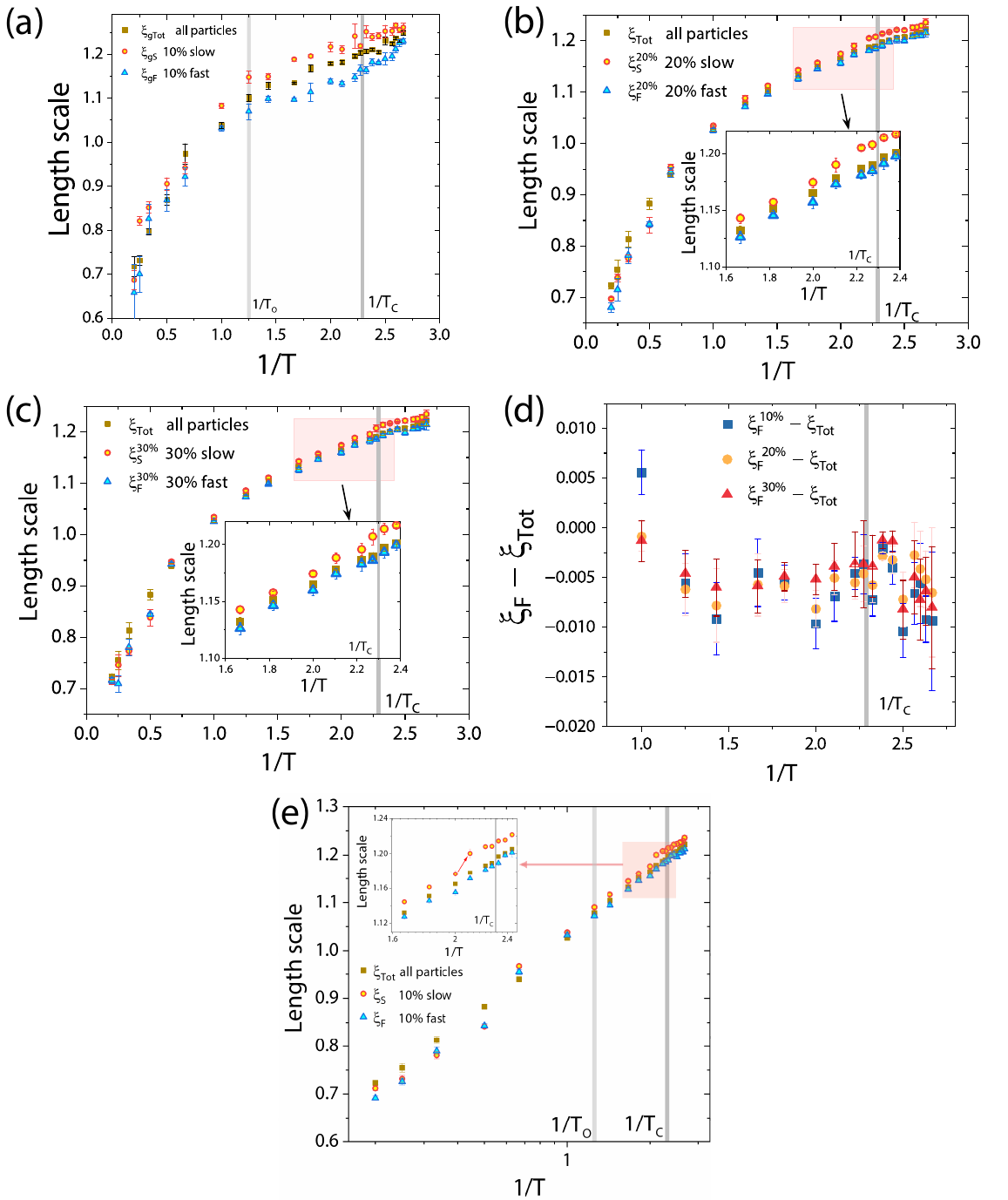}
\caption{Temperature dependence of length scales associated with dynamical heterogeneity. (a) Length scales extracted from the decay of $g(r)$ for all particles ($\xi_{gTot}$), top 10\% slow particles ($\xi_{gS}$) and top 10\% fast particles ($\xi_{gF}$). (b) and  (c) Length scale from the decay of $S_\rho(6, r)$ for 20\% and 30\% thresholds, respectively. The insets show a zoom of the main panel near $T_c$. (d) Difference of the length scales of fast particles and all particles for different mobility cut-offs. (e) The length scale $\xi_{\rm Tot}$, $\xi_S$, $\xi_F$ as a function of $\log(1/T)$ shows a linear dependence for $\xi_{\rm Tot}$ and $\xi_F$ at intermediate and low $T$. Inset: Zoom on the $T$-range close to $T_c$, showing the jump in $\xi_S$.\\
}
\label{SI_fig:5}
\end{figure}

\noindent In Fig.~\ref{SI_fig:5}(a), we show the length scale extracted from $g(r)$ for all particles ($\xi_{gTot}$) along with those associated with dynamical heterogeneity ($S$ and $F$ particles) for a 10\% mobility cut-off. As expected, the characteristic length scales of the slow particles ($\xi_{gS}$) exceed those of the fast particles ($\xi_{gF}$), and all length scales grow with cooling. However, due to the insufficient SNR of the $g(r)$ signal, we do not observe any remarkable $T-$dependence of these length scales at any temperature. Thus, this observable fails to detect the structural changes associated with the $S$ particles that the $S_\rho$ analysis readily picks up at $T_c$ (Fig. 1(d) of the main text and Figs.~\ref{SI_fig:5}(b) and (c)).

Figure 1(d) of the main text shows the length scales for $S$ and $F$ particles for a mobility cut-off of 10\%. In Figs.~\ref{SI_fig:5}(b) and (c), we show the corresponding data for 20\% and 30\% mobility cut-offs.
For all cut-off values, we observe a step in $\xi_S$ near $T_c$ while no such feature is found in $\xi_F$. This is also evident in the difference $\xi_F-\xi_{Tot}$, which shows no discernible peak-like feature near $T_c$ (Fig.~\ref{SI_fig:5}(d)). 

\section*{Quantifying slow particle clusters}

\begin{figure}
\centering
\includegraphics[width=1.0\textwidth]{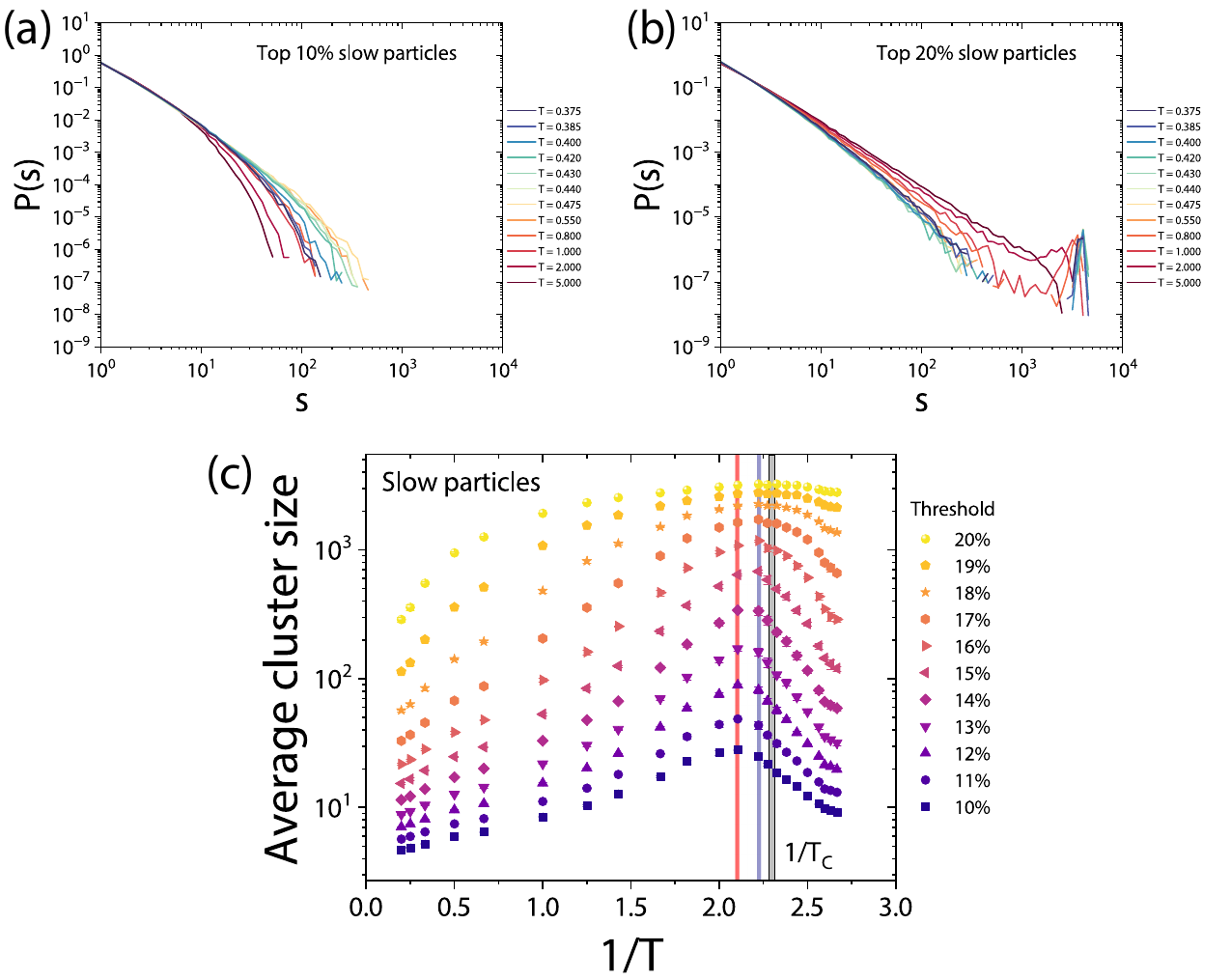}
\caption{Cluster size distribution of slow particles for different mobility cut-offs. Distribution of slow particle clusters based on a nearest-neighbor distance cut-off of $1.4\sigma$ and mobility cut-off of $10\%$ (a) and $20\%$ (b). (c) Temperature dependence of average cluster size for different mobility cut-offs between 10\% and 20\%. The red and blue vertical lines near $T_c$ denote the position of peaks for 10\% and 20\% mobility cut-offs, respectively.}
\label{SI_fig:6}
\end{figure}

\noindent Figures~\ref{SI_fig:6}(a) and (b) show the cluster-size distribution, $P(s)$, of slow particles for 10\% and 20\% mobility 
cut-offs, respectively. Two $S$ particles belong to the same cluster if they are within $1.4\sigma$ of each other. Upon cooling, we observe that $P(s)$ for the 20\% mobility cut-off shows a clear peak at very large cluster sizes. This feature strongly suggests a percolation-like transition of $S$ particles.

In Fig.~\ref{SI_fig:6}(c), we show the temperature dependence of the average cluster size, defined as $\sum s^2P(s))/\sum sP(s)$. For mobility cut-offs 
$ \gtrapprox 10$\%, the temperature 
at which the 
average cluster size is maximal coincides with the temperature at which
we observed a step-like increase 
in $\xi_S^{(10\%)}$ (Fig. 1(d) of main text). As the mobility threshold is increased towards 20\%, this peak gradually shifts closer to $T_c$, consistent with the trend in $\xi_S^{(20\%)}$. Taken together, these observations support a percolation-like scenario of $S$ particles near $T_c$.  

\section*{Hopping dependent directional correlation function}

\noindent We define the hopping direction as the vector $\bf{r}$ connecting the position of a particle at $t=0$ and $t=t^*$. For fast particles, this vector points in the direction of cage escape, whereas for the slow particles, which only rattle within their cages, this vector points in a random direction and serves as a reference. 

\begin{figure}
\centering
\includegraphics[width=0.8\textwidth]{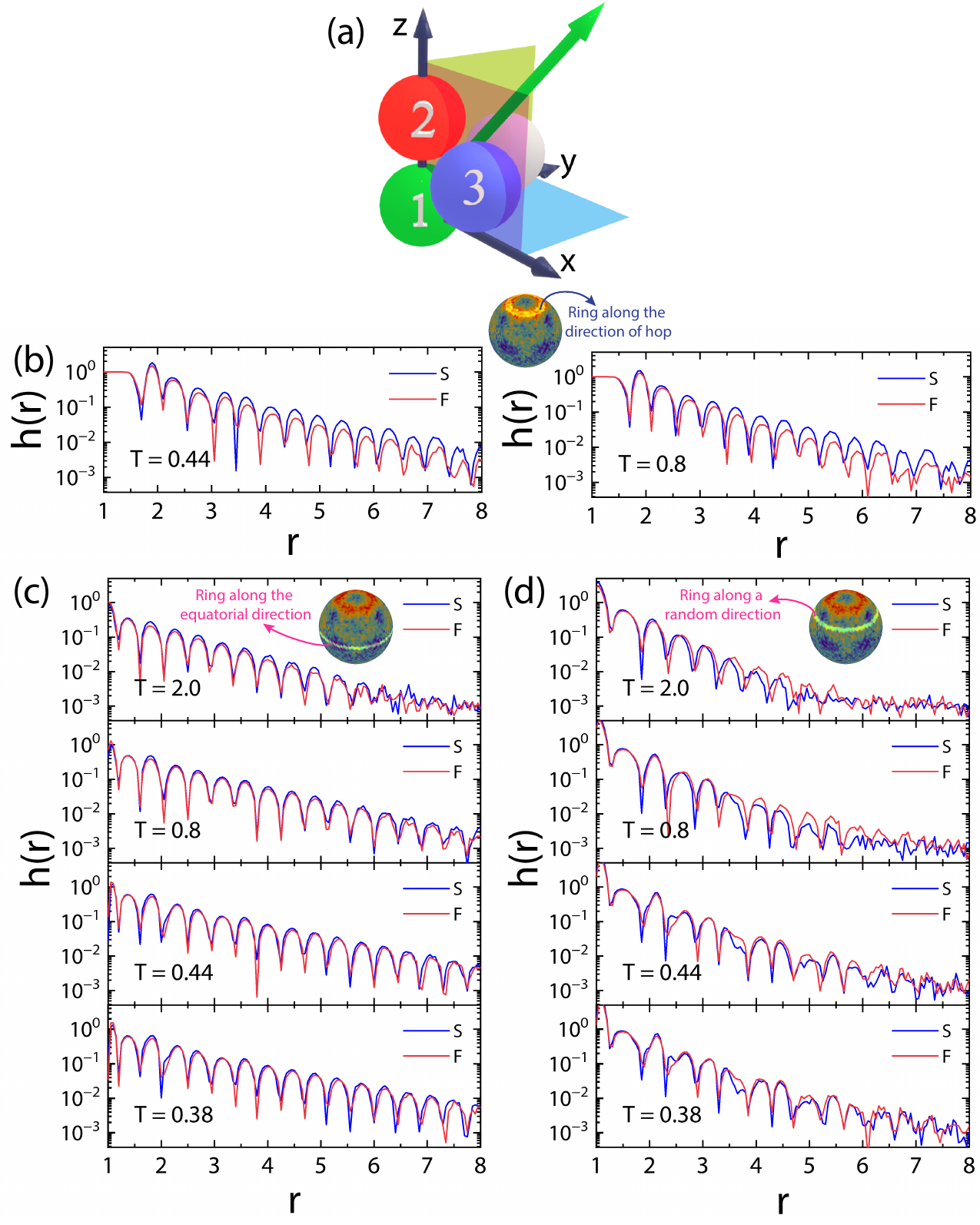}
\caption{Correlation along hopping direction. (a) Locally oriented coordinate system with the hopping direction in the first octant. (b) Directional correlation function in the LCS, $h(r) = |g(r)-1|$, (illustrated in Fig. 2(a) of the main manuscript)  along the hopping direction for $T = 0.44$ an $T = 0.8$. The region over which these correlations are evaluated are indicated by rings.
(c) and (d) Directional correlation along equatorial (c) and a random direction (d) for different temperatures.
}
\label{SI_fig:7}
\end{figure}

Figure~\ref{SI_fig:7}(a) shows the local coordinate system (LCS) in which the hop direction lies in the first octant. The construction of this LCS involves placing the particle of interest ($\#1$) at the origin. Next, the particle in the nearest-neighbor shell of particle $\#1$  and closest to the hopping direction ($\#2$) is placed along the $z-$axis. The final constraint requires selection of a particle $\#3$ (a mutual neighbour of particles $\#1$ and $\#2$), which, when placed in the $x-z$ plane, constrains the hopping direction
to be in the first octant. This triplet forms the LCS that has been used to
obtain the density fields shown in Fig. 2(c-e) of the main text. To improve the statistics while evaluating the directional correlation, we relax the final constraint. Now the hopping vector is not constrained to be in the first octant, but instead traces a cone. We calculate the directional correlation function, $h({\bf r}) = |g({\bf r})-1|$, over all vectors $\bf r$ that are inside this conical region, i.e., have an opening angle between $23^\circ$ and $33^\circ$ with the $z-$axis (Fig.~\ref{SI_fig:7}(b) middle panel). In Fig. 3(a) of the main text, we show $h({\bf r})$ for $S$ and $F$ particles for $T = 2.0$ and $T = 0.38$. In Fig.~\ref{SI_fig:7}(b), we present this correlation function for two other temperatures, $T = 0.44$ and $T = 0.8$. We again see that structural order around the slow particles is more pronounced than the fast ones. To demonstrate that the difference in structure between slow and fast particles is primarily along the cage escape direction, we calculated $h({\bf r})$ for $S$ and $F$ particles with $\bf r$ lying within a thin equatorial slice (Fig. \ref{SI_fig:7}(c)) as well as along a random direction (Fig.~\ref{SI_fig:7}(d)). For both these cases, $h({\bf r})$ for slow and fast particles is nearly identical proving that it is the weak structural ordering along the cage escape direction that governs particle mobility. 

\begin{figure}
\centering
\includegraphics[width=1.0\textwidth]{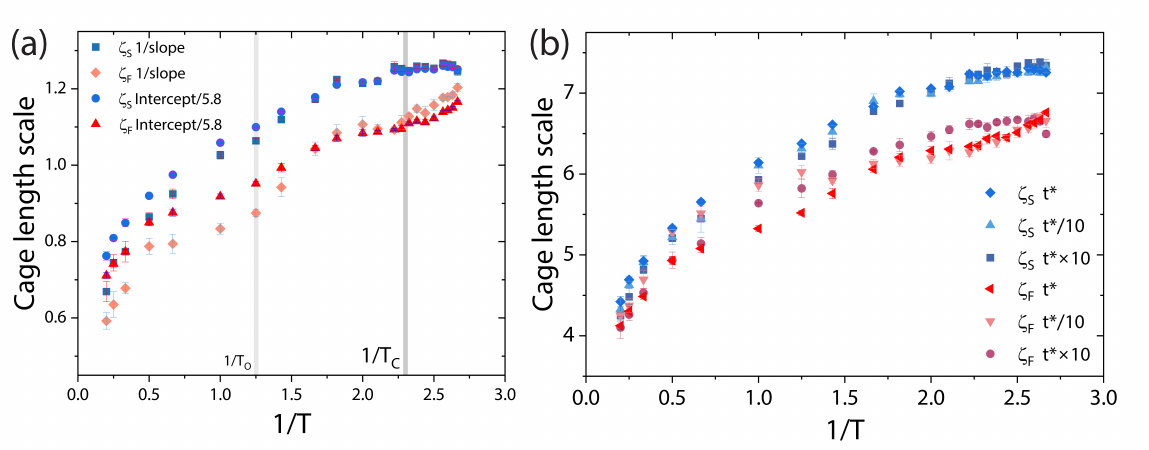}
\caption{Cage length scale. (a) Cage length scale extracted using two different procedures, i.e.,  based on the slope or the intercept of $h(r)$. (b) Dependence of the length scale on the timescale used for classifying slow and fast particles. 
}
\label{SI_fig:8}
\end{figure}

The cage length scales $\zeta_S$ and $\zeta_F$ can be extracted either as a slope of the decay of the directional correlation function $h({\bf r})$ or when this function has fallen below a fixed threshold value (operationally set at $10^{-2}$). In Fig.~\ref{SI_fig:8}(a), we show that the striking $T-$dependence seen in these length scales, especially $\zeta_F$, is insensitive to the procedure used to obtain the length scales i.e., the approach is robust. 

In Fig. \ref{SI_fig:8}(b), we show how the $T-$dependence of $\zeta_S$ and $\zeta_F$ depends on the timescales used to classify particles as $S$ and $F$. While for both $t = t^*/10$ and $t = 10t^*$, we find that $\zeta_S>\zeta_F$, like for $t = t^*$, the plateaus in $\zeta_F$ observed near $T_o$ and $T_c$ are most evident for $t = t^*$. This can be understood as follows: For $t = 10t^*$, the fast particle has a non-negligible probability to have broken the cage many times, and since these escape directions are not all aligned along $\bf r$, the cage structural anisotropy is averaged out, and the two plateaus vanish (see dark brown circles). For $t = t^*/10$, the fast particles have yet to sample the cage structural anisotropy completely, and again the plateaus in $\zeta_F$ are absent (light brown triangles). As expected, this sensitive dependence on the choice of time scale used to classify particles is weaker for the $S$ particles, since they do not break out of their cages and $\zeta_s^{t^*/10}\approx \zeta_s^{t^*} \approx \zeta_s^{10t^*}$ at all temperatures.\\

\newpage

\newpage

\newpage

\newpage

\newpage

\newpage

\newpage

\begin{figure*}[h]
\centering
\includegraphics[width=0.8\textwidth]{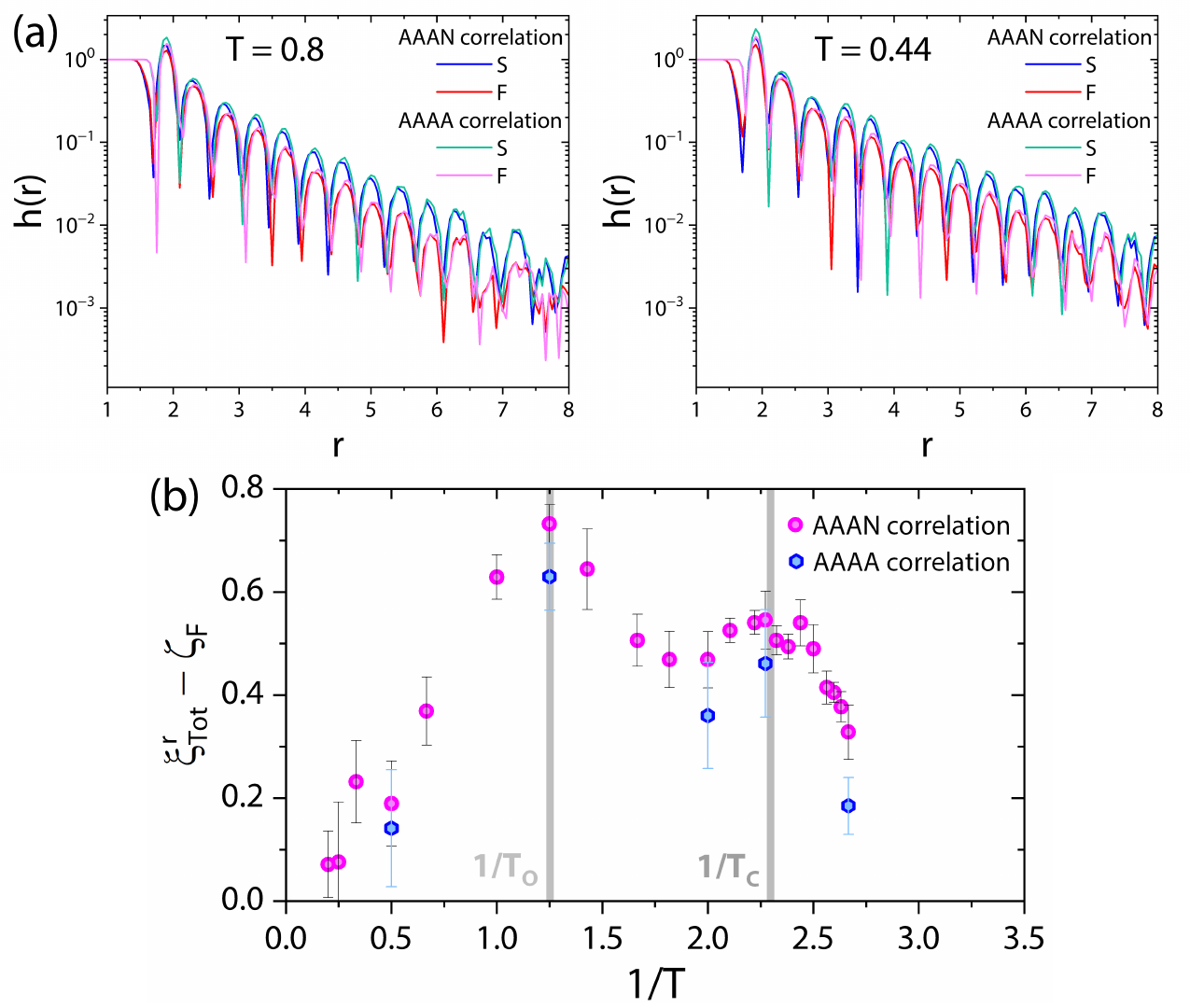}
\caption{Comparison between AAAN and AAAA correlations. (a) The spatial correlation function $h(r)$ at two temperatures. The curves show the correlation function if the fourth particle is of any type and of type A. One notes that, independent of $T$, the AAAA correlation is a bit higher than the one for AAAN, and this holds for the fast as well as the slow particles. (b) Cage asymmetry parameter $\epsilon = \xi^{r}_{\rm Tot} - \zeta_F$ with $\zeta_F$ obtained from AAAA correlation (blue) for 5 temperatures. 
Magenta: Data from Fig.~3(c) of the manuscript.}
\label{fig:9}
\end{figure*}

\clearpage 

\end{document}